\documentstyle[12pt]{article}

\catcode`\@=11
\def\makepreprititle{\par
  \begingroup
  \def\thefootnote{\fnsymbol{footnote}}
  \def\
@makefnmark{\hbox
  to 0pt{$^{\@thefnmark}$\hss}}
  \if@twocolumn
  \twocolumn[\@makepreprititle]
  \else \newpage
  \global\@topnum\z@
  \@makepreprititle \fi\thispagestyle{empty}\@thanks
  \endgroup
  \setcounter{footnote}{0}
  \let\makepreprititle\relax
  \let\@makepreprititle\relax
  \gdef\@thanks{}\gdef\@author{}\gdef\@title{}
  \gdef\@preprintnumber{}\gdef\@preprintdate{}\gdef\subtitle{}
  \let\thanks\relax}
\def\preprintnumber#1{\gdef\@preprintnumber{#1}}
\def\preprintdate#1{\gdef\@preprintdate{#1}}
\def\subtitle#1{\gdef\@subtitle{#1}}
\def\@makepreprititle{\newpage
{\def\baselinestretch{1}
  \begin{flushright} \@preprintnumber \par
  \@preprintdate \end{flushright} } \par
  \begin{center}
\vskip 1.5em
  {\LARGE \@title \par} \vskip 2.5em
  {\large \lineskip .5em
  \begin{tabular}[t]{c}\@author
  \end{tabular}\par}
  \vskip 1em {\large \@date} \end{center}
  \par
  \vfil}
\date{\sl Department of Physics, Tohoku University\\Sendai, 980 Japan}
\preprintdate{~}
\preprintnumber{~}
\subtitle{}
\def\abstract{\if@twocolumn
\section*{Abstract}
\else \normalsize
\begin{center}
{\bf Abstract\vspace{-.5em}\vspace{0pt}}
\end{center}
\quotation
\addtocounter{page}{-1}
\fi}
\def\endabstract{\if@twocolumn\else\endquotation\fi}
\def\spacing#1{\def\baselinestretch{#1}
\typeout{baselinestretch is modified to \baselinestretch}}
\catcode`\@=12

\spacing{1.5}
\renewcommand{\thefootnote}{\fnsymbol{footnote}}
\makeatletter
\long\def\@makefntext#1{\parindent 1em\noindent
 \hbox to 1.8em{\hss$[{\@thefnmark}]$ }#1}
\def\@makefnmark{\hbox{$^{[{\@thefnmark}]}$}}
\makeatother
\preprintnumber{TU-464}
\preprintdate{Sep., 1994}
\title{ Naturally Light Higgs Doublets in the Supersymmetric Grand
Unified Theories with Dynamical Symmetry Breaking}
\author{ T.Yanagida\\
\large \sl Department of Physics, Tohoku University\\
\large \sl Sendai, 980 Japan}
\date{~}
\begin{document}
\makepreprititle
\begin{abstract}
We construct a supersymmetric grand unified model in which a dynamical
symmetry breaking of SU(5)$_{GUT}$ generates large masses for
color-triplet Higgs multiplets while keeping SU(2)$_L$-doublet Higgs
multiplets naturally light. The vanishing masses for the Higgs doublets
are guaranteed by the Peccei-Quinn symmetry. We
also show that the Peccei-Quinn symmetry suppresses the dangerous dimension 5
operators for the nucleon decay.
\end{abstract}
\renewcommand{\thefootnote}{\arabic{footnote}}
\setcounter{footnote}{0}
\newpage
A basic assumption of the grand unified theories (GUT's) \cite{GUT} is
the presence of a large hierarchy between two mass scales $M_{GUT}\sim
10^{16}$GeV and $m_W\sim 10^2$GeV. It is,
however, very much unlikely that such a large mass hierarchy survives
the quantum correction without any symmetry reason. Supersymmetry (SUSY)
is well known symmetry \cite{SUSY} to protect the hierarchy built at the
tree level against the radiative corrections. However, SUSY itself is
unable to explain the tree-level hierarchy. In fact, one has to fine
tune unrelated parameters in the superpotential to realize the required
large hierarchy in the minimum SUSY-GUT \cite{SUSY-GUT}. Although there
have been proposed several mechanisms \cite{MNTY,Ant,BB} to build
naturally the hierarchy in SUSY-GUT's, each mechanism has its own
problem  and no convincing solution has been found so far.

The purpose of this letter is to show a new solution to this serious
problem in the SUSY-GUT's. We find that a dynamical symmetry breaking at
the GUT scale produces large masses for color-triplet Higgs multiplets
while SU(2)$_L$-doublet Higgs multiplets are kept light naturally. Here,
the vanishing masses for the Higgs doublets are guaranteed by the
Peccei-Quinn (PQ) symmetry \cite{PQ}. We note
that the dangerous dimension 5 ($d=5$) operators \cite{SYW}
for the nucleon decay are suppressed by this PQ symmetry.

The model is based on a strong hypercolor gauge group SU(3)$_H$ acting
on 7+7 chiral supermultiplets,
\begin{eqnarray}
\phi_\alpha^A = \varphi_\alpha^A + \theta\psi_\alpha^A,\nonumber\\
\bar{\phi}_A^\alpha = \bar{\varphi}_A^\alpha +
\theta\bar{\psi}_A^\alpha,\\
(\alpha=1-3,      ~ A=1-7) \nonumber
\end{eqnarray}
which transform as ${\bf 3}$ + ${\bf 3}^\ast$ under the SU(3)$_H$. At
the classical level this theory has a global flavor symmetry,
\begin{eqnarray}
G_0 = SU(7)_1\times SU(7)_2\times U(1)_1\times U(1)_2\times U(1)_R,
\end{eqnarray}
where U(1)$_R$ corresponds to the phase rotation of the SU(3)$_H$ gauge
fermions ($R$ symmetry). The axial U(1)$_A$ $\equiv$ U(1)$_{1-2}$ and U(1)$_R$
have
hypercolor SU(3)$_H$ anomalies, with a linear combination
U(1)$_{\bar{R}}$ of these two symmetries being anomaly free. Thus, the
global continuous symmetry at the quantum level is
\begin{eqnarray}
G &=& SU(7)_1\times SU(7)_2\times U(1)_V\times U(1)_{\bar{R}},
\end{eqnarray}
where the U(1)$_V$ is a diagonal subgroup U(1)$_{1+2}$.

We make a dynamical assumption that the Higgs and confining phases are
smoothly connected \cite{complem}. This complementarity picture has been
observed by lattice calculations \cite{complem} in the theories with
elementary scalar fields being the fundamental representations of the
gauge group. Since the scalar components $\varphi_\alpha^A$ and
$\bar{\varphi}_A^{\alpha}$ in the present model are all the fundamental
${\bf 3}$ and ${\bf 3}^\ast$ of SU(3)$_H$, our dynamical assumption
seems quite reasonable \cite{BPY}. Using this complementarity picture,
we first discuss a vacuum (i.e. symmetry breaking) in the Higgs phase and
then reinterpret the result in the confining phase of SU(3)$_H$.

In the Higgs phase, the scalar fields $\varphi_\alpha^A$ and
$\bar{\varphi}_A^\alpha$ have vacuum-expectation values of the form,
\begin{eqnarray}
\label{Higgsvacuum}
\langle\varphi_\alpha^A\rangle = \langle\bar{\varphi}_A^\alpha\rangle =
v\delta_A^{\alpha},\\
(\alpha=1-3).\nonumber
\end{eqnarray}
In this vacuum the global symmetry G$_0$ is spontaneously broken down to
\begin{eqnarray}
\label{breaking}
G_0 \longrightarrow H_0 = SU(3)_C\times SU(4)_1\times SU(4)_2\times
U(1)'_1\times U(1)'_2\times U(1)'_R.
\end{eqnarray}
In the confining phase, the symmetry breaking Eq.~(\ref{breaking}) in the
Higgs phase is achieved by the following SU(3)$_H$ invariant condensation;
\begin{eqnarray}
\label{condens}
 \langle\phi_\alpha^A\bar{\phi}_B^\alpha\rangle = \left\{
\begin{array}{rl}
v^2\delta_B^A,&\quad\mbox{{\rm for} $A,B=1-3$}\\
 0, &{{\rm for~others}}
\end{array}\right.
\end{eqnarray}
\begin{eqnarray}
\label{condens2}
\epsilon^{\alpha\beta\gamma}\epsilon_{ABC}\langle\phi_\alpha^A\phi_\beta^B\phi_\gamma^C\rangle &=& \epsilon_{\alpha\beta\gamma}\epsilon^{ABC}\langle\bar{\phi}_A^\alpha\bar{\phi}_B^\beta\bar{\phi}_C^\gamma\rangle = v^3, ~~~~~~~~{\rm for} ~ A,B,C=1-3
\end{eqnarray}
where $\epsilon_{\alpha\beta\gamma}, \epsilon_{ABC},....$ are the
third-rank antisymmetric tensors. This condensation is, therefore, a
basic assumption in the present analysis.

We  now gauge SU(5)$_{GUT}$$\times$U(1)$_S$ which is a subgroup of the global
symmetry G$_0$. The elementary chiral multiplets, $\phi_\alpha^I$ and
$\bar{\phi}_I^\alpha~ (I=1-5)$, are assumed to transform as ${\bf 5}$
and ${\bf 5}^\ast$ under the grand unified SU(5)$_{GUT}$ and the
multiplets, $\phi_\alpha^{\ell+5}$  and $\bar{\phi}_{\ell
+5}^\alpha~(\ell=1,2)$, are singlets of
the SU(5)$_{GUT}$.( $I,J$ denote the SU(5)$_{GUT}$ indices which run
from 1 to 5 while $A,B,C$ the global SU(7)$_{1,2}$ indices running from
1 to 7.) Charges of the U(1)$_S$ for
$\phi_\alpha^A$ and $\bar{\phi}_A^\alpha$ are chosen as
$Q_S(\phi_\alpha^A) = 1$ and $Q_S(\bar{\phi}_A^\alpha) =-1$ for all
$A=1-7$.

The condensation given in Eqs.~(\ref{condens},\ref{condens2}) causes
the breaking of the
flavor gauge symmetry,
\begin{eqnarray}
SU(5)_{GUT}\times U(1)_S \longrightarrow SU(3)_C\times SU(2)_L\times
U(1)_Y,
\end{eqnarray}
where U(1)$_Y$ is a linear combination of an U(1) subgroup of
SU(5)$_{GUT}$ and the U(1)$_S$.\footnote{ The necessity of introducing
the extra gauge group U(1)$_S$ comes from that the condensation in
Eq.~(\ref{condens2}) breaks an U(1) subgroup of the SU(5)$_{GUT}$. If one
assumes only the condensation in Eq.~(\ref{condens}), one has the
desired breaking of SU(5)$_{GUT}$, SU(5)$_{GUT}\longrightarrow $
SU(3)$_C\times$SU(2)$_L\times$U(1)$_Y$. If it is the case, one needs not
introduce the extra U(1)$_S$ and the unification of three gauge
coupling constants, $g_C, g_L$ and $g_Y$, is an automatic result of the
grand unification. To check whether this vacuum in fact exists an intensive
study on
the dynamics of SUSY confining forces are required, that is, however,
beyond the scope of the present paper.} \footnote{The
U(1)$_Y$ gauge multiplet V$_Y$ is given by
\begin{eqnarray}
V_Y = \frac{\left(\sqrt{15}g_{1S}V_{24} +
g_5V_S\right)}{\sqrt{15g_{1S}^2+g_5^2}} \nonumber
\end{eqnarray}
where V$_{24}$ and V$_S$ are gauge vector multiplets of the U(1)
subgroup of SU(5)$_{GUT}$ and the extra U(1)$_S$, respectively. The
gauge multiplet orthogonal to the V$_Y$ receives a large mass from the
condensation in Eq.~(\ref{condens2}). However, in the
limit $g_{1S}\rightarrow \infty$, the V$_Y$ is dominated by the V$_{24}$.
Thus, one may understand that the GUT relation in Eq.~(\ref{relation})
holds in this strong coupling limit as shown in the text. As for the
definition of $g_5$ and V$_{24}$ see Ref.\cite{HMY} for example, and for the
coupling constant $g_{1S}$
see below in the text.} We assume that all quark, lepton and
Higgs multiplets have vanishing charges of U(1)$_S$ and hence they
belong to the standard representations of SU(5)$_{GUT}$. Then, the
coupling constant $g_Y$ of the U(1)$_Y$ gauge multiplet to
the ordinary fields is given by
\begin{eqnarray}
g_Y = g_5 (1 + \frac{\alpha_5}{15\alpha_{1S}} )^{-1/2}
\end{eqnarray}
where $\alpha_5 \equiv g_5^2/4\pi$ and $\alpha_{1S}
\equiv g_{1S}^2/4\pi$ are the gauge coupling constants of the
SU(5)$_{GUT}$ and U(1)$_S$ at the GUT scale v, respectively. Here, the
coupling constant $g_S$ is defined as
\begin{eqnarray}
L_{int} = \int d^4\theta \left\{\phi_\alpha^{*A}e^{-g_SV_S}\phi_\alpha^A +
\bar{\phi}_A^{*\alpha}e^{g_SV_S}\bar{\phi}_A^\alpha \right\}
\end{eqnarray}
with V$_S$ being the gauge vector multiplet of the U(1)$_S$.

In the strong coupling limit of the U(1)$_S$, i.e. $\alpha_{1S} \rightarrow
\infty$, we recover the GUT unification of the three gauge coupling
constants \footnote{ The flavor gauge symmetry and its breaking is very
similar to those in the flipped SU(5) model \cite{Ant}. However, in the
flipped model one needs a fine tuning between the two gauge coupling
constants of SU(5) and U(1) in order to have the GUT relation in
Eq.~(\ref{relation}).}
\begin{eqnarray}
\label{relation}
g_C = g_L = g_Y = g_5.
\end{eqnarray}
If one requires this GUT relation by $1\%$ accuracy, one gets  a
constraint, $\alpha_{1S} \geq 0.2$, for $\alpha_5 \simeq
1/20$ at the GUT scale.\footnote{ Since the U(1)$_S$ is not asymptotic free
theory, the
coupling $\alpha_{1S}$ may blow up below the Planck scale. Thus, it is very
interesting to embed the strong gauge groups SU(3)$_H$ and U(1)$_S$ into
some larger non-Abelian group together. This possibility will be investigated
elsewhere. Another solution to this problem is to consider that the
strong U(1)$_S$ has an ultra-violet fixed point.}

Let us discuss massless bound states in the confining phase, switching
off the flavor gauge interactions (i.e. $g_5=g_{1S}=0$). Corresponding to the
symmetry breaking $G_0 \longrightarrow H_0$ in Eq.~(\ref{breaking}), 58
mssless composite Nambu-Goldstone(NG) bosons appear. They are scalar
components of the massless composite chiral multiplets
\cite{buch},\footnote{ In the Higgs phase we have the corresponding NG
chiral multiplets,
\begin{eqnarray}
\bar{\phi}_{i+3}^\alpha,~~\phi_\alpha^{i+3},~~(\phi_\alpha^{b}+\bar{\phi}_{b}^\alpha),~~~(\phi_\alpha^{\alpha}-\bar{\phi}_{\alpha}^\alpha). \nonumber
\end{eqnarray}}
\begin{eqnarray}
\label{NGmultiplet}
\bar{\Phi}_{i+3}^a = (\phi_\alpha^{a}\bar{\phi}_{i+3}^\alpha),~~~~~~~~~~~
\Phi_a^{i+3} = (\phi_\alpha^{i+3}\bar{\phi}_{a}^\alpha),\nonumber\\
\Phi_b^a = (\phi_\alpha^{a}\bar{\phi}_{b}^\alpha),~~~~~~~~~~~~~~~~~\\
\Phi_0 =
\epsilon_{\alpha\beta\gamma}\epsilon_{abc}(\phi_\alpha^{a}\phi_\beta^{b}\phi_\gamma^{c} - \bar{\phi}_{a}^\alpha\bar{\phi}_{b}^\beta\bar{\phi}_{c}^\gamma),\nonumber  \\
(a,b,c=1-3,~i=1-4).  \nonumber
\end{eqnarray}
Notice that $a,b,c=1-3$ and $i+3=4,5$ denote the color SU(3)$_C$ and
SU(2)$_L$ indices, but $\alpha,\beta,\gamma=1-3$ the hypercolor
SU(3)$_H$ indices. The $\Phi_b^a$ and $\Phi_0$ contain 9 and 1 quasi-NG
bosons,\cite{buch,BPY}, respectively. Thus, we have totally 68 massless
scalar bosons.

The unbroken group H$_0$ has two U(1)'s. The new $R$ symmetry U(1)$'$$_R$ is
free from the hypercolor anomaly. The axial U(1)$'$$_A$
$\equiv$U(1)$'$$_{1-2}$ is broken down to a discrete Z$_8$ by the SU(3)$_H$
anomaly. Thus, the unbroken symmetry H at the quantum level is
\begin{eqnarray}
\label{unbrokenH}
H = SU(3)_C\times SU(4)_1\times SU(4)_2\times U(1)'_V\times
U(1)'_R\times Z_8,
\end{eqnarray}
where U(1)$'$$_V$ is a diagonal subgroup of U(1)$'$$_1$
$\times$U(1)$'$$_2$, i.e.U(1)$'$$_V$ $\equiv$U(1)$'$$_{1+2}$.

With the unbroken H in Eq.~(\ref{unbrokenH}) we have checked that the chiral
fermions of the
composite NG multiplets in Eq.~(\ref{NGmultiplet}) satisfy all 'tHooft
anomaly matching conditions\cite{tHooft}. This also strongly supports
our dynamical assumption in Eq.~(\ref{condens},\ref{condens2}).

When we switch on the SU(5)$_{GUT}$$\times$ U(1)$_S$ gauge interactions,
12 NG chiral multiplets, $\Phi_{i+3}^a$ and $\bar{\Phi}_a^{i+3}$ ($i=1,2$) and
one $\Phi_0$ are absorbed to massive vector multiplets of
SU(5)$_{GUT}$$\times$ U(1)$_S$. To eliminate unwanted 9 $\Phi_b^a$ in
the masslss spectrum one may introduce a mass for $\phi_\alpha^I({\bf 5})$
and $\bar{\phi}_I^\alpha({\bf 5}^\ast)$, $m_\phi$
$\phi_\alpha^I\bar{\phi}_I^\alpha$. However, in this case the
condensation $\langle\phi_\alpha^I\rangle =
\langle\bar{\phi}_I^\alpha\rangle \neq 0$ breaks the SUSY in the Higgs
phase, since it gives a non-vanishing vacuum energy.
A simple way to solve this problem is to add a new chiral multiplet
$\Sigma_J^I$ transforming as (${\bf 1}+{\bf 24}$) representation of
the SU(5)$_{GUT}$. Then, we take the following superpotential;
\begin{eqnarray}
\label{sigmacoupling}
W = m_\phi\phi_\alpha^I\bar{\phi}_I^\alpha +
\lambda\phi_\alpha^I\Sigma_I^J\bar{\phi}_J^\alpha +
\frac{M_\Sigma}{2}Tr(\Sigma_J^I)^2, \\
(I,J=1-5).\nonumber
\end{eqnarray}
We have a SUSY-invariant vacuum in the Higgs phase,
\begin{eqnarray}
\langle\Sigma_J^I\rangle = \left\{
\begin{array}{rl}
-\frac{m_\phi}{\lambda}\delta_J^I, &\quad\mbox{{\rm for}~$I,J=1-3$} \\
 0, &\quad\mbox{{\rm for~others}}
\end{array}\right.
\end{eqnarray}
and $\langle\phi_\alpha^I\rangle$ and $\langle\bar{\phi}_I^\alpha\rangle$ given
in Eq.~(\ref{Higgsvacuum}) with
\begin{eqnarray}
v = \frac{1}{\lambda}\sqrt{m_\phi M_\Sigma}.
\end{eqnarray}

In the confining phase the Yukawa coupling,
$\lambda\phi_\alpha^I\Sigma_I^J\bar{\phi}_J^\alpha$, in
Eq.~(\ref{sigmacoupling}) induces a mass mixing between the $\Phi_b^a$ and
$\Sigma_J^I$ as
\begin{eqnarray}
W_{eff} \sim \lambda v\Sigma_{a}^{b}\Phi_b^a .
\end{eqnarray}
Diagonalization of the mass matrix for $\Phi$ and $\Sigma$ generates the mass
for the $\Phi_b^a$,\footnote{In the Higgs phase, the corresponding NG
multiplets $\frac{1}{\sqrt{2}}(\phi_\alpha^{a} + \bar{\phi}_{a}^\alpha)$
acquire also masses $\sim (\lambda v)^2/M_\Sigma$ through the mixing with
$\Sigma_J^I$.}
\begin{eqnarray}
W_{mass} \sim \frac{(\lambda v)^
2}{M_\Sigma} \Phi_b^a\Phi_a^b.
\end{eqnarray}

Composite states remaining in the massless spectrum are now
\begin{eqnarray}
\label{colorphi}
\Phi_a^{\ell +5} ~~~~{\rm and}~~~~\bar{\Phi}_{\ell +5}^a,~~~~(\ell =1,2)
\end{eqnarray}
which transform as ${\bf 3}$ and ${\bf 3}^\ast$ under the color
SU(3)$_C$.\footnote{If one assume
SU(2) as the hypercolor gauge group instead of the SU(3)$_H$, the
remaining massless states are SU(2)$_L$-doublets, which may be
identified as two pairs of Higgs doublets. In this case, however, one
needs nonrenormalizable interactions to have effective Yukawa couplings
of these Higgs multiplets to the ordinary quarks and leptons.} It
should be noted here that there still remains a continuous global
symmetry, SU(2)$_1\times$ SU(2)$_2\times$ U$(1)''_V$, which is a
subgroup of the SU(4)$_1\times$ SU(4)$_2\times$ U$(1)'_V$ in
Eq.~(\ref{unbrokenH}). The presence of the above two pairs ($\ell=1,2$)
of massless chiral multiplets is required by this unbroken chiral
SU(2)$_1\times$ SU(2)$_2$, where the composite $\Phi_a^{\ell+5}$ and
$\bar{\Phi}_{\ell+5}^a$ are ({\bf 2,1}) and ({\bf 1,2})
representations, respectively. There is also an axial U$(1)''_A$
under which the original chiral multiplets transform as
\begin{eqnarray}
\phi_\alpha^I,~\bar{\phi}_I^\alpha~ \longrightarrow
\phi_\alpha^I,~\bar{\phi}_I^\alpha,~~~(I=1-5) \nonumber \\
\phi_\alpha^{\ell+5},\bar{\phi}_{\ell+5}^\alpha \longrightarrow
e^{i\alpha}\phi_\alpha^{\ell+5}, e^{i\alpha}\bar{\phi}_{\ell+5}^\alpha
{}~~~(\ell=1,2).
\end{eqnarray}
Although this U$(1)''_A$ is broken down to a discrete Z$_4$ by the
hypercolor SU(3)$_H$ anomaly, the Z$_4$ keeps the masslessness of
the color-triplet composite supermultiplets in
Eq.~(\ref{colorphi}).\footnote{If one assumes 6+6 chiral
multiplets, $\phi_\alpha^A$ and $\bar{\phi}_A^\alpha~(A=1-6)$, one
have a pair of massless states, $\Phi_a^6$ and $\bar{\Phi}_6^a$, whose
mass seems to be forbidden by an axial U(1)$_A$ symmetry. However, the
U(1)$_A$ is broken down to Z$_2$ by the hypercolor anomaly and the mass
for these composite multiplets may be induced by the hypercolor
instanton effects. This is a main reason why we assume the hypercolor
theory with 7+7 chiral multiplets.}

It is now clear that we can easily incorporate the missing partner
mechanism \cite{MNTY} in the present model. Let us introduce two pairs of Higgs
multiplets $H_I^{(\ell)}$ and
$\bar{H}_{(\ell)}^I$ with $\ell=1,2$ that are ${\bf 5}$ and ${\bf 5}^\ast$ of
SU(5)$_{GUT}$ and add a possible Yukawa couplings,
\begin{eqnarray}
\label{HYukawa}
W = \bar{h}\phi_\alpha^I\bar{\phi}_{\ell+5}^\alpha H_I^{(\ell)} +
h\phi_\alpha^{\ell+5}\bar{\phi}_I^\alpha\bar{H}_{(\ell)}^I.
\end{eqnarray}
These Yukawa interactions preserve the chiral $SU(2)_1\times SU(2)_2$
symmetry under which the elementary chiral multiplets,
$\phi_\alpha^{\ell+2}, \bar{H}_{(\ell)}^I$ and
$\bar{\phi}_{\ell+5}^\alpha, H_I^{(\ell)}$ transform as ({\bf 2,1})
and ({\bf 1,2}) representations, respectively.

The Yukawa couplings in Eq.~(\ref{HYukawa}) give rise to masses for the
NG multiplets $\Phi_a^{\ell+5}$, $\bar{\Phi}_{\ell+5}^a$ in
Eq.~(\ref{colorphi}) and the elementary $H_{a}^{(\ell)},
\bar{H}_{(\ell)}^{a}$ as
\begin{eqnarray}
W_{mass} \sim \bar{h}v\bar{\Phi}_{\ell+5}^aH_{a}^{(\ell)} +
hv\Phi_a^{\ell+5}\bar{H}_{(\ell)}^{a}.
\end{eqnarray}
The SU(2)$_L$-doublet Higgs $H_i^{(\ell)}$ and $\bar{H}_{(\ell)}^i~
(i=4,5)$ remain naturally massless, since they do not have partners to
form invariant masses and, furthermore, the masslessness of these
doublet-Higgs multiplets is guaranteed by the unbroken SU(2)$_1\times$
SU(2)$_2$ symmetry.  On the other hand, the color-triplet
$H_{a}^{(\ell)}$ and $\bar{H}_{(\ell)}^{a}$ have the partners
$\bar{\Phi}_{\ell+5}^{a}$ and $\Phi_a^{\ell+5}$ and hence they can
form two pairs of massive scalar multiplets, whose masses are
invariant under the SU(2)$_1\times$SU(2)$_2$. It should be stressed
that the remaining unbroken symmetry is now
SU(3)$_{C}\times$SU(2)$_L\times$U(1)$_Y\times$SU(2)$_1\times$
SU(2)$_2\times$U(1$)''_V$.\footnote{The U(1$)'_R$ is explicitely broken
by the superpotential in Eq.~(\ref{sigmacoupling}).}, with which all
the 'tHooft anomaly matching conditions are satisfied. Namely, the
global anomalies arising from loops of the elementary fields,
$\phi_\alpha^A,~ \bar{\phi}_A^\alpha,~ H_I^{(\ell)}$ and $\bar{H}_{(\ell)}^I$,
are reproduced by the surviving doublet Higgs,
$H_i^{(\ell)}$ and $\bar{H}_{(\ell)}^i$.\footnote{When one examines
the chiral anomalies at the elementary-field level one should remember
that the generator of U(1)$_Y$ is a linear combination of generators
of the $24^{th}$ component of SU(5)$_{GUT}$ and of the U(1)$_S$.}

The first pair of the SU(2)$_L$-doublet Higgs $H_I^{(1)}$ and $\bar{H}_{(1)}^I$
are assumed to have Yukawa couplings to quark and  lepton chiral multiplets
${\bf 5}^\ast$ and ${\bf 10}$,
\begin{eqnarray}
\label{HY}
{\bf 5}^\ast\cdot{\bf 10}\bar{H}_{(1)}^I + {\bf 10}\cdot{\bf 10}H_I^{(1)}.
\end{eqnarray}


These Yukawa interactions breaks the chiral SU(2)$_1\times$ SU(2)$_2$,
but preserve the maximum subgroup U(1)$_1\times$U(1)$_2$. Important for the
present analysis is
the axial U(1)$_A\equiv$U(1)$_{1-2}$ under which
\begin{eqnarray}
H_I^{(1)},~\bar{H}_{(1)}^I\longrightarrow
e^{i\beta}H_I^{(1)},~e^{i\beta}\bar{H}_{(1)}^I, \nonumber \\
H_I^{(2)},~\bar{H}_{(2)}^I\longrightarrow
e^{-i\beta}H_I^{(2)},~e^{-i\beta}\bar{H}_{(2)}^I,\\
{\bf 5}^\ast,~{\bf 10}\longrightarrow e^{-(i/2)\beta}{\bf
5}\ast,~e^{-(i/2)\beta}{\bf 10}. \nonumber
\end{eqnarray}
This is nothing but the Peccei-Quinn (PQ) symmetry\cite{PQ}. This PQ
symmetry has no hypercolor SU(3)$_H$ anomly, but has a color SU(3)$_C$
anomaly providing a solution to the strong $CP$ problem. Thanks to
this U(1)$_A$ the two pairs of Higgs doublets, $H_i^{(\ell)}$ and
$\bar{H}_{(\ell)}^i~(i=4,5)$ still remain massless.\footnote{
It seems that the mass terms, $H_i^{(1)}\bar{H}_{(2)}^i$ and
$H_i^{(2)}\bar{H}_{(1)}^i$, are allowed. However, it is not the case,
since there is, in fact, a larger chiral symmetry
U(1)$_1\times$U(1)$_2$ as noted in the text. By this chiral
symmetry those mass terms are completely forbidden. The PQ U(1)$_A$ is
the diagonal subgroup U(1)$_{1-2}$ and another combination
U(1)$_{1+2}$ is so-called 5-ness symmetry defined as the following
transformations;
 \begin{eqnarray}
 \phi_\alpha^I,~~\bar{\phi}_I^\alpha \longrightarrow
\phi_\alpha^I,~\bar{\phi}_I^\alpha, \nonumber \\
 \phi_\alpha^6 \longrightarrow
e^{-2i\gamma}\phi_\alpha^6,~~~\bar{\phi}_6^\alpha \longrightarrow
e^{2i\gamma}\bar{\phi}_6^\alpha, \nonumber \\
\phi_\alpha^7 \longrightarrow e^{2i\gamma}\phi_\alpha^7,~~~\bar{\phi}_7^\alpha
\longrightarrow e^{-2i\gamma}\bar{\phi}_7^\alpha, \nonumber \\
 H_I^{(1)}\longrightarrow e^{-2i\gamma}H_I^{(1)},~~\bar{H}_{(1)}^I
\longrightarrow e^{2i\gamma}\bar{H}_{(1)}^I, \nonumber \\
 H_I^{(2)}\longrightarrow e^{2i\gamma}H_I^{(2)},~~\bar{H}_{(2)}^I
\longrightarrow e^{-2i\gamma}\bar{H}_{(2)}^I, \nonumber \\
{\bf 5}^\ast \longrightarrow e^{-3i\gamma}{\bf 5}^\ast,~~~
 {\bf 10} \longrightarrow e^{i\gamma}{\bf 10}. \nonumber
 \end{eqnarray}}

As for the breaking of the PQ symmetry we assume an asymmetric pair of
PQ multiplets\cite{MSY}, $\phi_P$ and $\phi_Q$, which transform as
$\phi_P\rightarrow e^{2i\beta}\phi_P$ and $\phi_Q\rightarrow
e^{-6i\beta}\phi_Q$. Since the $\phi_P$ can couple only to the second
pair of Higgs
$H_I^{(2)}$ and $\bar{H}_{(2)}^I$, the PQ breaking generates a mass at
O($10^{12}$) GeV for
$H_i^{(2)}$ and $\bar{H}_{(2)}^i$ while the first pair of Higgs
doublets $H_i^{(1)}$ and $\bar{H}_{(1)}^i$ with $i=1,2$
remain massless\cite{MSY}. As pointed out in Ref.~\cite{HMY2}, the
unification of three gauge coupling constants can be maintained as
soon as the masses of the color-triplet Higgs multiplets are taken as
$\sim$O($10^{14}$) GeV.

The final comment is on the nucleon decay. In the present model, the dangerous
$d=5$ operators\cite{SYW}, ${\bf
5}^\ast\cdot{\bf 10}\cdot{\bf 10}\cdot{\bf 10}$, are forbidden by the
PQ symmetry. Even if the mass term, $m_{(2)}H_I^{(2)}\bar{H}_{(2)}^I$, is
generated by the PQ breaking, the $d=5$ operators are still
suppressed, since $H_I^{(2)}$ and $\bar{H}_{(2)}^I$ do not have Yukawa
couplings to quark and lepton multiplets. Therefore, it is very
much unlikely that the $d=5$ operators give dominant contributions to the
nucleon decay.

\end{document}